\documentclass[useAMS,usegraphicx,usenatbib]{mn2e}          

\usepackage{times}

\newcommand{\Msun}{\ensuremath{\,{\rm M}_\odot}}                  
\newcommand{\Rsun}{\ensuremath{\,{\rm R}_\odot}}                  
\newcommand{\Teff}{\ensuremath{T_{\rm eff}}}                      
\newcommand{\Mjup}{\ensuremath{\,{\rm M}_{\rm Jup}}}              
\newcommand{\Rjup}{\ensuremath{\,{\rm R}_{\rm Jup}}}              
\newcommand{\Teq}{\ensuremath{T_{\rm eq}^{\,\prime}}}             
\newcommand{\safronov}{\ensuremath{\Theta}}                       
\newcommand{\ms}{\,m\,s$^{-1}$}                                   
\newcommand{\as}{\ensuremath{^{\prime\prime}}}                    
\newcommand{\FeH}{\ensuremath{\left[\frac{\rm Fe}{\rm H}\right]}} 
\newcommand{\Porb}{\ensuremath{P_{\rm orb}}}                      
\newcommand{\pjup}{\ensuremath{\,\rho_{\rm Jup}}}                 
\newcommand{\psun}{\ensuremath{\,\rho_\odot}}                     
\newcommand{\chir}{\ensuremath{\chi_\nu^{\,2}}}                   
\newcommand{\mc}[1]{\multicolumn{2}{c}{#1}}
\newcommand{\mcc}[1]{\multicolumn{3}{c}{#1}}
\newcommand{\er}[3]{\ensuremath{#1^{+#2}_{-#3}}}

\newcommand{\ermcc}[5]{\mcc{\ensuremath{{#1\,^{+#2}_{-#3}}\,^{+#4}_{-#5}}}}
\newcommand{\reff}[1]{{#1}}                                   

\title[Physical properties of HAT-P-13]
      {Refined physical properties of the HAT-P-13 planetary system}

\author[Southworth et al.]
       {John Southworth\,$^{1}$, I.\ Bruni\,$^{2}$, L.\ Mancini\,$^{3}$, J.\ Gregorio\,$^{4}$ \\
        $^{1}$\,Astrophysics Group, Keele University, Newcastle-under-Lyme, ST5 5BG, UK \\
        $^{2}$\,INAF -- Osservatorio Astronomico di Bologna, Via Ranzani 1, 40127 Bologna, Italy \\
        $^{3}$\,Max Planck Institute for Astronomy, K\"onigstuhl 17, 69117 -- Heidelberg, Germany \\
        $^{4}$\,Grupo Atalaia, CROW Observatory-Portalegre, Portugal
       }

\begin{document} \maketitle 

\begin{abstract}
We present photometry of four transits of the planetary system HAT-P-13, obtained using defocussed telescopes. We analyse these, plus nine datasets from the literature, in order to determine the physical properties of the system. The mass and radius of the star are $M_{\rm A} = 1.320\pm 0.048 \pm 0.039$\Msun\ and $R_{\rm A} = 1.756 \pm 0.043 \pm 0.017$\Rsun\ (statistical and systematic errorbars). We find the equivalent quantities for the transiting planet to be $M_{\rm b} = 0.906 \pm 0.024 \pm 0.018$\Mjup\ and $R_{\rm b} = 1.487 \pm 0.038 \pm 0.015$\Rjup, with an equilibrium temperature of $\Teq = 1725 \pm 31$\,K. Compared to previous results, which were based on much sparser photometric data, we find the star to be more massive and evolved, and the planet to be larger, hotter and more rarefied. The properties of the planet are not matched by standard models of irradiated gas giants. Its large radius anomaly is in line with the observation that the hottest planets are the most inflated, but at odds with the suggestion of inverse proportionality to the \FeH\ of the parent star. We assemble all available times of transit midpoint and determine a new linear ephemeris. Previous findings of transit timing variations in the HAT-P-13 system are shown to disagree with these measurements, and can be attributed to small-number statistics.
\end{abstract}



\section{Introduction}                                                                                                              \label{sec:intro}

The discovery of the HAT-P-13 by \citet{Bakos+09apj2} elicited substantial interest. The transiting extrasolar planet (TEP) HAT-P-13\,b and its host star HAT-P-13\,A are rather typical examples of these objects, with two exceptions. Firstly, the star is the second-most metal-rich known to host a TEP after XO-2\,A\footnote{See: {\tt http://www.astro.keele.ac.uk/$\sim$jkt/tepcat/}} \citep{Burke+07apj} (but see the discussion on host star \FeH\ values in \citealt{Enoch+11aj}). Secondly, there is a third component in the system which is clearly detected in the radial velocity measurements (RVs) of the host star \citep{Bakos+09apj2}. HAT-P-13\,c has an orbit with a period of $446.22 \pm 0.27$\,d, an eccentricity of $0.6616 \pm 0.0052$ and a minimum mass of $14.28 \pm 0.28$\Mjup\ \citep{Winn+10apj2}. This object is expected to induce transit timing variations (TTVs) within the HAT-P-13\,A,b system, which potentially allow the structure of the planet to be probed \citep{MardlingLin04apj,Batygin++09apj}. HAT-P-13 is unfortunately not the best system for such analyses, as its relatively long and shallow transits are not conducive to precise timing measurements.

An observation of the Rossiter-McLaughlin effect by \citet{Winn+10apj2} has shown that the projected angle between the orbital axis of HAT-P-13\,b and the rotational axis of the parent star is consistent with zero. This is in line with the fact that misaligned axes are only found for TEP systems containing a star hotter than roughly 6250\,K \citep{Winn+10apj3,Schlaufman10apj,Albrecht+11apj2}. \citet{Winn+10apj2} also found a long-term drift in the radial velocity measurements of the star, which may be the signature of a {\em fourth} component to the system on an orbit of much longer period.

\begin{table*} \centering
\caption{\label{tab:obslog} Log of the observations presented in this work. $N_{\rm obs}$ is the number
of observations and `Moon illum.' is the fractional illumination of the Moon at the midpoint of the transit.
The aperture sizes are the radii of the software apertures for the object, inner sky and outer sky, respectively.}
\begin{tabular}{llccccccccc} \hline
Transit    & Date & Start time & End time & $N_{\rm obs}$ & Exposure & Filter &    Airmass     & Moon   & Aperture   & Scatter \\
           &      &    (UT)    &   (UT)   &               & time (s) &        &                & illum. & sizes (px) & (mmag)  \\
\hline
Cassini    & 2011 02 06 & 19:21 & 00:37 & 127 & 120 & Thuan-Gunn $i$ & 1.17 $\to$ 1.00 $\to$ 1.10 & 0.129 & 25, 35, 55 & 0.77 \\
Cassini    & 2011 04 17 & 18:58 & 23:50 & 147 &  90 & Thuan-Gunn $i$ & 1.03 $\to$ 2.11            & 0.998 & 18, 30, 50 & 0.93 \\
Portalegre & 2011 01 31 & 22:13 & 05:21 & 128 & 180 & Cousins $I$    & 1.11 $\to$ 1.01 $\to$ 1.64 & 0.041 & 10, 35, 45 & 2.47 \\
Portalegre & 2011 02 03 & 20:53 & 02:55 & 119 & 150 & Cousins $I$    & 1.25 $\to$ 1.01 $\to$ 1.16 & 0.008 & 10, 35, 45 & 2.53 \\
\hline \end{tabular} \end{table*}

\citet{Szabo+10aa} attempted to detect a transit of the third body, at a time predicted by \citet{Winn+10apj2}, but were not successful. Their observational campaign included scrutiny of two transits of HAT-P-13\,b, whose times of occurrence agreed well with the predicted timings.
\citet{Pal+11mn} subsequently presented photometry of three transits, all of whose midpoints fell earlier than expected according to an ephemeris built on the observations of \citet{Bakos+09apj2} and \citet{Szabo+10aa}. \citeauthor{Pal+11mn} interpreted this as evidence of TTVs.
\citet{Nascimbeni+11aa2} presented high-speed photometry of five transits in early 2011, taken as part of the TASTE project \citep{Nascimbeni+11aa}. They confirmed that a linear ephemeris could not explain the available transit timings, and postulated that a sinusoidal TTV with an amplitude of 0.005\,d was a good match to the available transit timings. Sinusoidal TTVs have previously been seen for WASP-3 \citep{Pollacco+08mn,Maciejewski+10mn} and WASP-10 \citep{Christian+09mn,Maciejewski+11mn} but have not yet been confirmed\footnote{A sinusoidal TTV was claimed for OGLE-TR-111 by \citet{Diaz+08apj} but has been refuted by \citet{Adams+10apj}}.
The putative TTVs for HAT-P-13 have been challenged by \citet{Fulton+11aj}, who presented observations of ten transits over two observing seasons. They found that a linear ephemeris was an acceptable match to all transit timing measurements, with the exception of the first of the two obtained by \citet{Szabo+10aa}.

\begin{table} \centering \caption{\label{tab:lc} Excerpts of the light
curves of HAT-P-13. The full dataset will be made available at the CDS.}
\begin{tabular}{lcrr} \hline
Telescope  & BJD(TDB) & Diff.\ mag. & Uncertainty \\
\hline
Cassini    & 2455599.30664 &    0.00061 & 0.00090 \\
Cassini    & 2455599.52535 & $-$0.00085 & 0.00078 \\[2pt]
Cassini    & 2455669.29046 & $-$0.00014 & 0.00108 \\
Cassini    & 2455669.49355 &    0.00066 & 0.00117 \\[2pt]
Portalegre & 2455596.37076 & $-$0.5552  & 0.0018  \\
Portalegre & 2455596.62212 & $-$0.5460  & 0.0017  \\[2pt]
Portalegre & 2455593.42575 & $-$2.0886  & 0.0022  \\
Portalegre & 2455593.72313 & $-$2.0970  & 0.0024  \\
\hline \end{tabular} \end{table}

\begin{figure} \includegraphics[width=0.48\textwidth,angle=0]{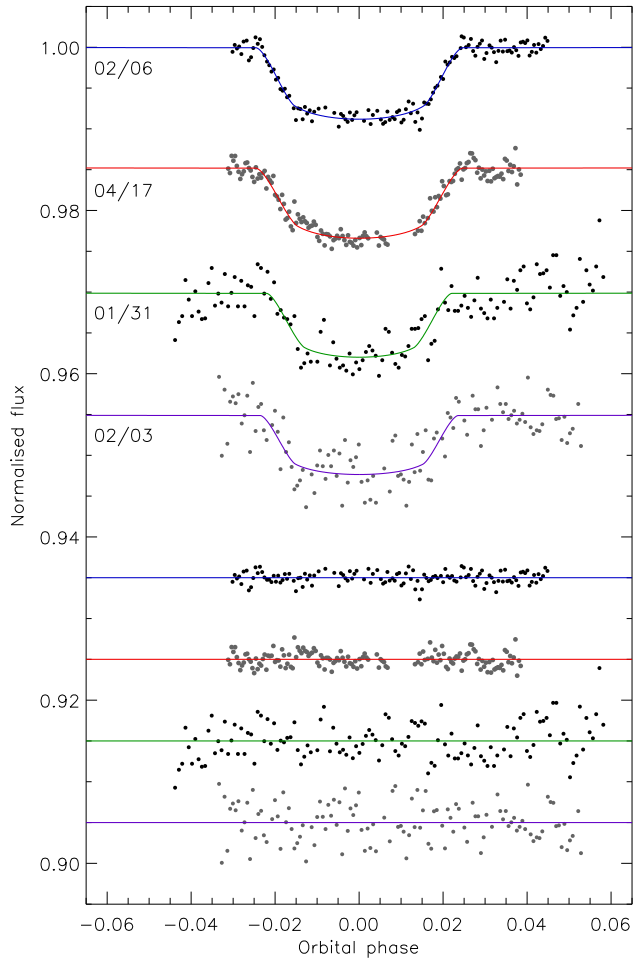}
\caption{\label{fig:lcfit1} New data presented in this work, compared to
the best {\sc jktebop} fits using the quadratic LD law. The dates of the
light curves are labelled using the format month/day. The residuals of the
fits are plotted in the lower half of the figure, offset from zero.} \end{figure}

The physical properties of the HAT-P-13 system have been derived by \citet{Bakos+09apj2} and \citet{Winn+10apj2}, who used the same light curves. Since these studies a wealth of new photometric data has been gathered. This has been used to investigate putative TTVs, but has not been brought to bear on improving the physical properties of the system. In this work we present new photometry covering four transits and use all available high-quality photometry to measure refined physical properties of HAT-P-13.


\section{Observations and data reduction}                                                                                             \label{sec:obs}

\begin{figure*} \includegraphics[width=\textwidth,angle=0]{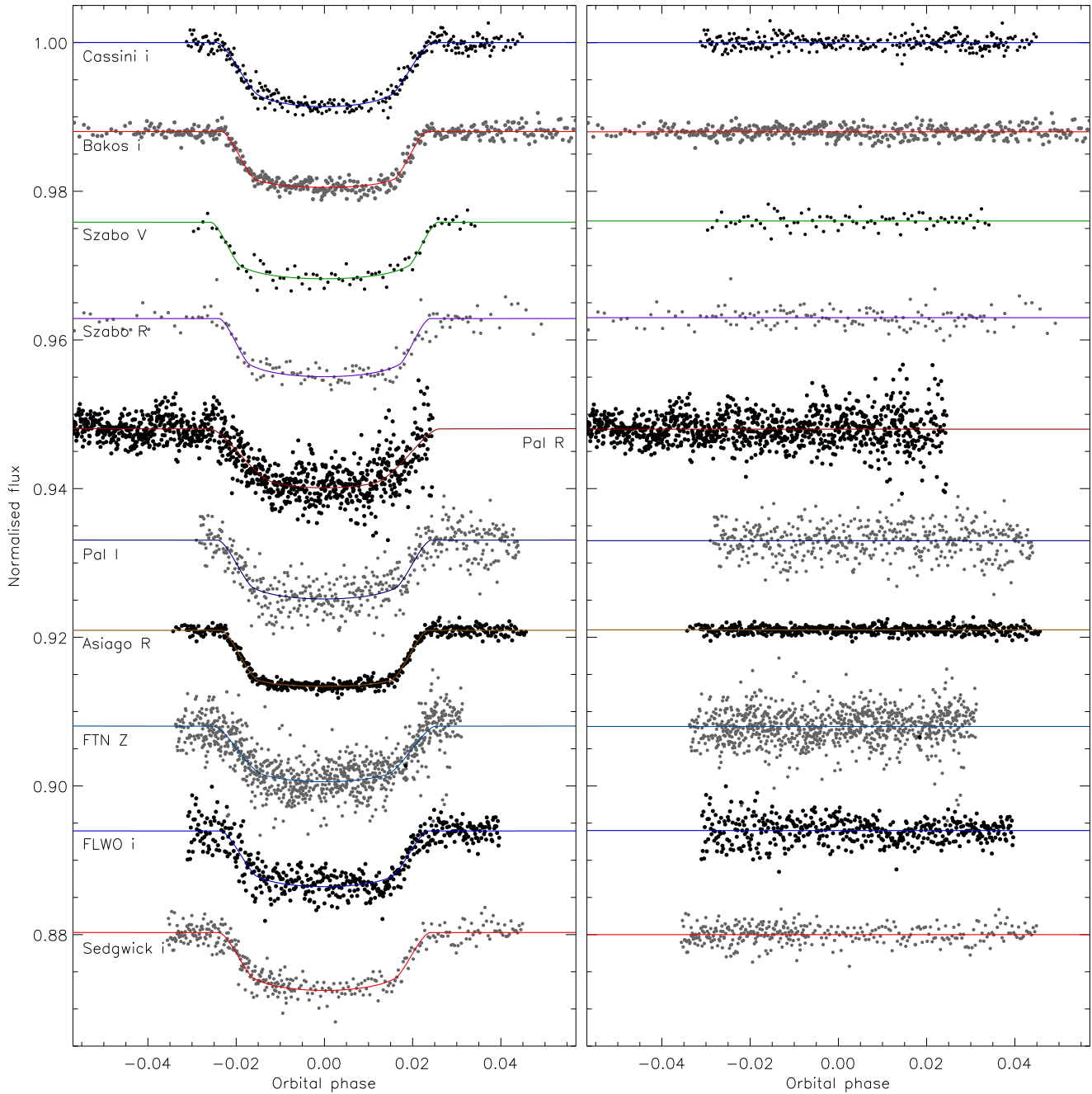}
\caption{\label{fig:lcfit2} Phased light curves of HAT-P-13 compared to the
best {\sc jktebop} fits using the quadratic LD law (left panel). They are shown
in the same order as in Table\,\ref{tab:lcfit}. The residuals of the fits are
plotted in the right panel, offset to bring them into the same relative position
as the corresponding best fit in the left panel.} \end{figure*}

\begin{table*} \caption{\label{tab:lcfit} Parameters of the {\sc jktebop}
fits to the light curves of HAT-P-13. The final parameters correspond to
the weighted mean of the results for the ten light curves.}
\begin{tabular}{l r@{\,$\pm$\,}l r@{\,$\pm$\,}l r@{\,$\pm$\,}l r@{\,$\pm$\,}l r@{\,$\pm$\,}l}
\hline
Source          & \mc{$r_{\rm A}+r_{\rm b}$} & \mc{$k$} & \mc{$i$ ($^\circ$)} & \mc{$r_{\rm A}$} & \mc{$r_{\rm b}$} \\
\hline
Cassini $i$-band         & 0.211  & 0.014  & 0.0932  & 0.0015  & 81.6  & 1.0  & 0.193  & 0.013  & 0.0180  & 0.0014  \\
Bakos FLWO $i$-band      & 0.193  & 0.011  & 0.08592 & 0.00084 & 82.61 & 0.87 & 0.178  & 0.010  & 0.0153  & 0.0010  \\
Szab\'o $V$-band         & 0.192  & 0.030  & 0.0836  & 0.0046  & 83.8  & 3.2  & 0.178  & 0.028  & 0.0148  & 0.0031  \\
Szab\'o $R$-band         & 0.207  & 0.024  & 0.0872  & 0.0026  & 81.7  & 1.8  & 0.190  & 0.022  & 0.0165  & 0.0019  \\
P\'al $R$-band           & 0.217  & 0.034  & 0.0895  & 0.0073  & 81.5  & 2.6  & 0.200  & 0.031  & 0.0178  & 0.0036  \\
P\'al $I$-band           & 0.210  & 0.019  & 0.0887  & 0.0032  & 81.5  & 1.3  & 0.193  & 0.017  & 0.0171  & 0.0018  \\
Nascimbeni $R$-band      & 0.2103 & 0.0048 & 0.08668 & 0.00052 & 81.97 & 0.36 & 0.1852 & 0.0044 & 0.01606 & 0.00042 \\
Fulton FTN $Z$-band      & 0.231  & 0.023  & 0.0882  & 0.0041  & 80.2  & 1.6  & 0.212  & 0.021  & 0.0187  & 0.0021  \\
Fulton FLWO $i$-band     & 0.208  & 0.016  & 0.0868  & 0.0023  & 81.5  & 1.2  & 0.191  & 0.015  & 0.0166  & 0.0017  \\
Fulton Sedgwick $i$-band & 0.199  & 0.017  & 0.0873  & 0.0029  & 82.4  & 1.4  & 0.183  & 0.016  & 0.0159  & 0.0017  \\
\hline
{\bf Final results} & \mc{ } & \mc{ } & {\bf 81.93} & {\bf 0.26} & {\bf 0.1863} & {\bf 0.0034} & {\bf 0.01622} & {\bf 0.00034} \\
\hline
\citet{Bakos+09apj2}     &   \mc{0.1856}   & 0.0844  & 0.0013  & 83.4  & 0.6  & 0.1712 & 0.0076 &   \mc{0.01445}    \\
\citet{Winn+10apj2}      &   \mc{0.1839}   & 0.08389 & 0.00081 & 83.40 & 0.68 & 0.1697 & 0.0072 &   \mc{0.01424}    \\
\citet{Fulton+11aj}      &   \mc{0.1967}   & 0.0855  & 0.0011  & 82.45 & 0.46 & 0.1812 & 0.0056 &   \mc{0.01549}    \\
\hline \end{tabular} \end{table*}

Two full transits of HAT-P-13 were observed with the BFOSC imager mounted on the 1.52\,m G.\ D.\ Cassini Telescope\footnote{Information on the 1.52\,m Cassini Telescope and BFOSC can be found at {\tt http://www.bo.astro.it/loiano/}} at Loiano Observatory, Italy. We used a Gunn $i$ filter and autoguided throughout. We had to reject a small number of datapoints in both transits, as they were affected by pointing jumps which compromised the data quality. A summary of our observations is given in Table\,\ref{tab:obslog} and the full data can be found in Table\,\ref{tab:lc}.

The telescope was defocussed so the point spread functions (PSFs) resembled annuli of widths 15--25 pixels, in order to reduce the light from the target and comparison stars to a maximum of roughly 35\,000 counts per pixel. This approach reduces the susceptibility of the data to flat-fielding noise and increases the efficiency of the observations. A detailed description of the defocussing method can be found in \citet{Me+09mn,Me+09mn2}, and an instance of its use with the Cassini telescope in \citet{Me+10mn}. Several images were taken with the telescope properly focussed, and used to verify that there were no faint stars within the defocussed PSF of HAT-P-13.

Data reduction was undertaken using standard methods pertaining to aperture photometry. Software aperture positions were specified by hand but shifted to account for pointing variations, which were found by cross-correlating each image against the reference image used to place the apertures. We found that the results were insensitive to the choice of aperture sizes (within reason) and to whether flat fields were used in the data reduction process. Differential-photometry light curves were obtained with respect to an optimal ensemble of four comparison stars constructed as outlined by \citet{Me+09mn}. The times of observation were converted to barycentric Julian date on the TDB timescale, using the {\sc idl} procedures of \citet{Eastman++10pasp}.

Two full transits were observed by JG from CROW-Portalegre, Portugal, using an f/5.6 30\,cm Schmidt-Cassegrain telescope, a KAF1603 CCD camera operating at a plate scale of 1.12\as\,px$^{-1}$, and a Cousins $I$ filter. The data were reduced by standard methods, using median-combined bias, dark and flat-field calibration observations. Aperture photometry was performed with C-Munipack\footnote{\tt http://c-munipack.sourceforge.net/} and differential-magnitude light curves obtained with respect to an ensemble of five comparison stars.


\section{Light curve analysis}                                                                                                   \label{sec:analysis}


We have analysed the Cassini and literature photometric observations of HAT-P-13 by the methods of the {\it Homogeneous Studies} project \citep{Me08mn,Me09mn,Me10mn,Me11mn}, which are briefly summarised below. The light curves and their best-fitting models are shown in Fig.\,\ref{fig:lcfit1} for the data presented in this paper, and in Fig.\,\ref{fig:lcfit2} for previously published observations. The ensuing parameters of the fit are given in Table\,\ref{tab:lcfit} and detailed results for each dataset can be found in an online-only supplement. The Portalegre transits were analysed using the same methods, but only the transit times are used below due to the comparatively large scatter of these data.

The light curves were modelled using the {\sc jktebop}\footnote{{\sc jktebop} is written in {\sc fortran77} and the source code is available at {\tt http://www.astro.keele.ac.uk/$\sim$jkt/}} code. The primary fitted parameters were the sum and ratio of the fractional radii of the star and planet, $r_{\rm A}${$+$}$r_{\rm b}$ and $k = \frac{r_{\rm b}}{r_{\rm A}}$, and the orbital inclination, $i$. The fractional radii of the components are defined as $r_{\rm A} = \frac{R_{\rm A}}{a}$ and $r_{\rm b} = \frac{R_{\rm b}}{a}$ where $a$ is the orbital semimajor axis, and $R_{\rm A}$ and $R_{\rm b}$ are the true radii of the two objects. \reff{Additional parameters of the fit included the magnitude level outside transit and the midpoint of the transit.}

We generated solutions with each of five limb darkening (LD) laws (linear, quadratic, square-root, logarithmic and cubic), and with three different treatments of the LD coefficients. The first possibility is to fix both coefficients to values predicted using model atmospheres; this leads to a dependence on stellar theory as well as slightly worse fits due to the larger number of degrees of freedom. The second option is to fit for both coefficients, but this is possible only when the data are of extremely high quality. Unless otherwise stated, we go for a third alternative: fit for the linear LD coefficients and fix the nonlinear one to theoretically predicted values (`LD-fit/fix' in the nomenclature of \citealt{Me10mn}). The two coefficients are highly correlated \citep[e.g.][]{Me++07aa} so the theoretical dependence inherent in this approach is negligible.

Uncertainties in each solution were calculated in two ways: from 1000 Monte Carlo (MC) simulations \citep{Me++04mn2}, and with a residual-permutation (RP) algorithm \citep{Jenkins++02apj}. The larger of the two possible errorbars was retained for each fitted parameter. Orbital eccentricity ($e$) and periastron longitude ($\omega$) were incorporated using the constraints $e\cos\omega = -0.0099 \pm 0.0036$ and $e\sin\omega = -0.0060 \pm 0.0069$ \citep{Winn+10apj2}. \reff{These constraints were treated as observational data and $e\cos\omega$ and $e\sin\omega$ were included as fitted parameters.}

\subsection{Analysis of each dataset}

The two Cassini transits were modelled together, after scaling the observational errors for each transit to give a reduced $\chi^2$ of $\chir = 1.0$. This step is necessary because the errors returned by the aperture photometry routine we use are usually too small. Due to the possibility of TTVs in the HAT-P-13 system one has to be careful when combining data. In this case we fitted for the orbital period (\Porb) and the midpoint of the first transit, which is equivalent to fitting for the two transit midpoints. The RP errorbars were selected because they are larger than the MC ones. The LD-fit/fix results were adopted. A final value for each photometric parameter was obtained by taking the weighted mean of the four values from the fits for the non-linear LD laws. Its errorbar was taken to be the largest of the four alternative values, with a contribution added in quadrature to account for any dependence of the parameter value on the choice of LD law.

\citet{Bakos+09apj2} presented $i$-band data obtained with the 1.2\,m telescope and KeplerCam at the F.\ L.\ Whipple Observatory (FLWO). The 3719 datapoints extend over seven transits within an interval of approximately one year, but only two of these transits have full phase coverage. We therefore converted the timestamps into orbital phase (using the ephemeris calculated by \citealt{Fulton+11aj}), sorted them and combined each set of eight consecutive points, to obtain 466 phase-binned points. This will wash out any TTVs present over that time interval, but inspection of fig.\,7 in \citet{Bakos+09apj2} shows that no significant variations exist. A preliminary fit returned $\chir = 7.53$ so the errorbars were scaled up by $\sqrt{7.53}$. The LD-fit/fix results are adopted and combined as above. The RP errors were larger than the MC ones, which is unusual for phase-binned data \citep{Me11mn} but accounted for in our analysis.

\citet{Szabo+10aa} obtained $V$- and $R$-band coverage of two transits; we did not consider the $V$-band data of the second transit as it suffers from systematic errors. The $R$-band observations were solved with \Porb\ as a fitted parameter, so the solutions are not sensitive to the effects of putative TTVs. Given the limited quantity of the data we did not attempt LD-fitted solutions. In both cases we found that correlated noise was unimportant and adopted the LD-fit/fix solutions. The $T_0$ values from the two datasets (referenced to cycle $-12$) are in poor agreement with each other (see Tables A3 and A4), and roughly bracket the time of midpoint derived by \citet{Szabo+10aa}. All three timings deviate from the ephemeris derived below by much more than the derived uncertainties.

\citet{Pal+11mn} presented photometry of three transits from two telescopes located at Konkoly Observatory. The first transit was obtained with the Schmidt telescope and a CCD camera equipped with a Bessell $I$ filter. The second and third came from the 1.0\,m telescope with VersArray CCD camera and a Cousins $R$ filter. The two datasets were modelled separately after scaling up their errorbars to enforce $\chir = 1.0$. For both bands we adopted the LD-fit/fix solutions. The MC errorbars are larger than the RP ones, indicating that red noise is not important.

The data presented by \citet{Nascimbeni+11aa2} were obtained with the Asiago 1.82\,m telescope and AFOSC imager, through a Cousins $R$ filter. They comprise 12\,585 observations covering five closely adjacent transits with cadences ranging from 5.8\,s to 9.7\,s. The data were taken over only 38 days, so can be combined without suffering smearing effects due to TTVs with periodicities above several months. We therefore phase-binned them by a factor of 25 into 504 bins, during which process a 4$\sigma$ clip removed 11 of the observations. A preliminary fit returned $\chir = 1.07$ so the supplied observational errors were left unmodified. The results show a strong preference for weaker LD than theoretical expectations, and the values for $r_{\rm A}$ and $r_{\rm b}$ depend somewhat on the treatment of LD. The LD-fixed solutions can be rejected due to a lower quality of fit, and the LD-fitted equivalents return unphysical coefficients, so the LD-fit/fix alternatives were adopted. RP errors are smaller than MC ones, as we ordinarily find for phase-binned data.

\citet{Fulton+11aj} obtained photometry of ten transits of HAT-P-13, of which only four were fully covered. These include two transits taken just over a year apart using the Faulkes Telescope North (FTN) and a $Z$ filter, which were solved with a free \Porb\ to allow for the possibility of TTVs. A third transit was obtained in the $i$ band with the FLWO 1.2\,m. Finally, one $i$-band transit from the Sedgwick 0.8\,m is accompanied by a partial transit obtained only three nights earlier, allowing these data to be modelled with \Porb\ fixed to any reasonable value. We found that the errorbars of the Sedgwick and FTN data had to be multiplied by $\sqrt{8.05}$ and $\sqrt{5.35}$, respectively, to obtain $\chir = 1.0$. For the FTN and Sedgwick data we were able to adopt the LD-fit/fix solutions, but for FLWO had to stick to LD-fixed as attempts to fit for LD coefficients returned unphysical values. For the FTN and FLWO data the RP errors are moderately larger than those from the MC algorithm, implying that correlated noise is significant in these light curves.

\begin{table} \begin{center}
\caption{\label{tab:minima} \reff{Times of minimum light of HAT-P-13
and their residuals versus the ephemeris derived in this work.}}
\setlength{\tabcolsep}{2.5pt}
\begin{tabular}{l@{\,$\pm$\,}l r r l} \hline
\multicolumn{2}{l}{Time of minimum}      & Cycle  & Residual & Reference \\
\multicolumn{2}{l}{BJD(TDB) $-$ 2400000} & no.    & (JD)     &           \\
\hline
54581.62443 & 0.00122 & $-$204.0 & $-$0.00183 & \citet{Bakos+09apj2}      \\   
54777.01324 & 0.00100 & $-$137.0 & $-$0.00099 & \citet{Bakos+09apj2}      \\   
54779.92990 & 0.00063 & $-$136.0 & $-$0.00057 & \citet{Bakos+09apj2}      \\   
54782.84394 & 0.00155 & $-$135.0 & $-$0.00277 & \citet{Bakos+09apj2}      \\   
54849.92099 & 0.00075 & $-$112.0 &    0.00080 & \citet{Bakos+09apj2}      \\   
54882.00078 & 0.00150 & $-$101.0 &    0.00197 & \citet{Bakos+09apj2}      \\   
54960.74005 & 0.00178 &  $-$74.0 &    0.00281 & \citet{Bakos+09apj2}      \\   
55167.79647 & 0.00280 &   $-$3.0 &    0.00631 & Gary(AXA)                 \\   
55194.03566 & 0.00229 &      6.0 & $-$0.00065 & \citet{Fulton+11aj}       \\   
55196.95450 & 0.00127 &      7.0 &    0.00195 & \citet{Fulton+11aj}       \\   
55199.86837 & 0.00123 &      8.0 & $-$0.00041 & Tieman(TRESCA)            \\   
55199.86867 & 0.00131 &      8.0 & $-$0.00011 & \citet{Fulton+11aj}       \\   
55231.94542 & 0.00091 &     19.0 & $-$0.00199 & \citet{Fulton+11aj}       \\   
55249.45117 & 0.00200 &     25.0 &    0.00634 & \citet{Szabo+10aa}        \\   
55269.86567 & 0.00180 &     32.0 &    0.00717 & Gary(AXA)                 \\   
55272.77577 & 0.00120 &     33.0 &    0.00103 & Gary(AXA)                 \\   
55272.77627 & 0.00250 &     33.0 &    0.00153 & Foote(AXA)                \\   
55275.69207 & 0.00180 &     34.0 &    0.00109 & Gary(AXA)                 \\   
55275.69312 & 0.00266 &     34.0 &    0.00214 & \citet{Fulton+11aj}       \\   
55307.77077 & 0.00370 &     45.0 &    0.00117 & Gary(AXA)                 \\   
55310.69197 & 0.00250 &     46.0 &    0.00613 & Gary(AXA)                 \\   
55511.90854 & 0.00141 &    115.0 &    0.00226 & \citet{Fulton+11aj}       \\   
55558.56302 & 0.00098 &    131.0 & $-$0.00307 & \citet{Pal+11mn}          \\   
55561.48416 & 0.00400 &    132.0 &    0.00183 & \citet{Pal+11mn}          \\   
55564.39876 & 0.00180 &    133.0 &    0.00019 & \citet{Nascimbeni+11aa2}  \\   
55584.81455 & 0.00153 &    140.0 &    0.00231 & Dvorak(TRESCA)            \\   
55590.64523 & 0.00179 &    142.0 &    0.00052 & \citet{Pal+11mn}          \\   
55593.55879 & 0.00185 &    143.0 & $-$0.00216 & This work (Portalegre)    \\   
55593.56147 & 0.00115 &    143.0 &    0.00052 & \citet{Nascimbeni+11aa2}  \\   
55596.47291 & 0.00140 &    144.0 & $-$0.00428 & Naves(TRESCA)             \\   
55596.47327 & 0.00202 &    144.0 & $-$0.00392 & This work (Portalegre)    \\   
55596.47662 & 0.00305 &    144.0 & $-$0.00057 & \citet{Nascimbeni+11aa2}  \\   
55599.39267 & 0.00075 &    145.0 & $-$0.00076 & \citet{Nascimbeni+11aa2}  \\   
55599.39446 & 0.00100 &    145.0 &    0.00103 & This work (Cassini)       \\   
55602.31068 & 0.00167 &    146.0 &    0.00101 & \citet{Nascimbeni+11aa2}  \\   
55613.97390 & 0.00225 &    150.0 & $-$0.00072 & \citet{Fulton+11aj}       \\   
55616.89290 & 0.00152 &    151.0 &    0.00204 & \citet{Fulton+11aj}       \\   
55619.80786 & 0.00134 &    152.0 &    0.00076 & \citet{Fulton+11aj}       \\   
55622.72351 & 0.00166 &    153.0 &    0.00018 & \citet{Fulton+11aj}       \\   
55669.38140 & 0.00126 &    169.0 & $-$0.00175 & This work (Cassini)       \\   
\hline \end{tabular} \end{center} \end{table}

\subsection{Combined results}

The photometric parameters resulting from the ten light curves are collected in Table\,\ref{tab:lcfit}. We calculated the weighted means to obtain final values of these quantities. The \chir\ values of the averaging process are all below 0.5, with the exception of $k$. We deduce that the ten light curves are in sufficient mutual agreement. The final values are dominated by the results from the Asiago light curve, which is easily the best of the available datasets.

For $k$ the averaging process yields $\chir=2.0$, so the agreement between the light curve solutions is not so good. The Cassini data are the primary contributor to this situation, as they point to a larger $k$ than the rest of the datasets. Individual solutions of the two Cassini light curves yield similar values of $k$. Moderate disagreements in $k$ have been frequently seen in the {\it Homogeneous Studies} papers and can be attributed to starspot activity and/or systematic errors in the light curve. $k$ is also determined to a high precision (by comparison to $r_{\rm A}$, $r_{\rm b}$ and $i$) so systematic differences are comparatively obvious.

Our final photometric parameters (Table\,\ref{tab:lcfit}) are somewhat different to values based on the discovery photometry \citep{Bakos+09apj2,Winn+10apj2}. In particular we find a lower $i$ and larger $r_{\rm A}$. The latter quantity is observationally strongly tied to $r_{\rm b}$ so our results point towards a larger star and planet than previously proposed. Our photometric parameters agree reasonably well with those proposed recently by \citet{Fulton+11aj}, but have smaller uncertainties.

\section{On the transit timings of HAT-P-13\,b}                                                                                      \label{sec:porb}

\begin{figure*} \includegraphics[width=\textwidth,angle=0]{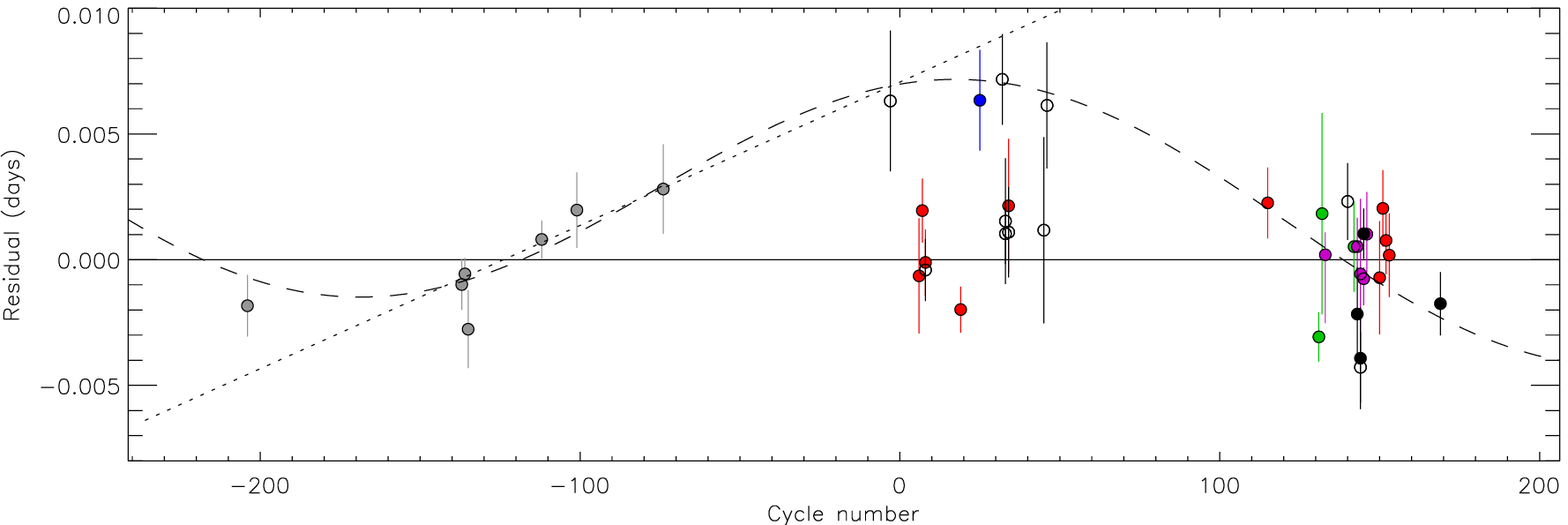}
\caption{\label{fig:minima} \reff{Plot of the residuals of the timings of mid-transit of
HAT-P-13 versus a linear ephemeris. The timings in black are from this work, in grey are
from \citet{Bakos+09apj2}, blue from \citet{Szabo+10aa}, lilac from \citet{Pal+11mn}, green
from \citet{Nascimbeni+11aa2}, red from \citet{Fulton+11aj}, and open circles for the amateur
timings. The solid line shows the ephemeris from the current work and the dotted line that
from \citet{Bakos+09apj2}. The dashed curve is an approximate representation of the possible
periodicity proposed by \citet{Nascimbeni+11aa2}.}} \end{figure*}

The third body in the HAT-P-13 system, known to have an eccentric orbit with $e_{\rm c} = 0.6616 \pm 0.0052$ and ${\Porb}_{\rm,c} = 446.22 \pm 0.27$\,d from radial velocity measurements, is expected to cause TTVs within the inner HAT-P-13\,A,b system \citep{Bakos+09apj2,PayneFord11apj}. \citeauthor{Bakos+09apj2} found ``suggestive'' but not ``significant'' evidence for TTVs in their data obtained in their 2007-8 and 2008-9 observing seasons. \citet{Szabo+10aa} obtained an improved orbital ephemeris and placed an upper limit of 0.001\,d on size of the phenomenon, with the inclusion of their two transit times from the 2009-10 season.

However, the times of three transits acquired by \citet{Pal+11mn} during the 2010-11 season were about 22 minutes early with respect to the previous orbital ephemerides, leading \citeauthor{Pal+11mn} to claim a significant detection of TTVs. The timings found by \citeauthor{Pal+11mn} were supported by \citet{Nascimbeni+11aa2} on the basis of five new transits obtained during the same observing season. \citeauthor{Nascimbeni+11aa2} demonstrated that a sinusoidal TTV function provided a good fit to all existing timing measurements.

\citet{Fulton+11aj} have subsequently presented ten transit timings which cast doubt on the possibility of TTVs: five from the 2010-11 season which agree well with those from \citet{Pal+11mn} and \citet{Nascimbeni+11aa2}, and five from the previous season which conflict with the timings found by \citet{Szabo+10aa}. \citeauthor{Fulton+11aj} reanalysed all published follow-up observations of HAT-P-13 and concluded that they were consistent with a linear ephemeris with the exception of the first transit dataset from \citet{Szabo+10aa}.

In order to firmly establish the character of the situation, we have collected all available transit midpoint times for the HAT-P-13\,A,b system. We have used the timings as quoted by the original sources, rather than adopting those from the reanalysis by \citet{Fulton+11aj}. We included ten timings obtained by amateur astronomers and placed on the AXA\footnote{Amateur Exoplanet Archive, {\tt http://brucegary.net/AXA/x.htm}} and TRESCA\footnote{The TRansiting ExoplanetS and CAndidates (TRESCA) website can be found at, {\tt http://var2.astro.cz/EN/tresca/index.php}} websites. We rejected amateur timings which are based on data that are either very scattered or do not cover a full transit.

\reff{Transit timings were obtained from our own observations by fitting {\sc jktebop} models to each transit following the LD-fit/fix prescription. The flux normalisation was allowed to vary linearly with time. Errors were estimated from RP and from 1000 MC simulations, and were multiplied by two in order to guard against any undetected systematic noise in the data. At the request of the referee we also assessed correlated noise using the `$\beta$' approach \citep[e.g.][]{Winn+07aj2}. We evaluated values for individual transits and for groups of between two and ten datapoints, finding a maximum $\beta$ of $1.26$. The corresponding increases in the uncertainties in the $T_0$ values are smaller than the factor of two we used above.}

All timings were placed on the BJD(TDB) time system. \reff{We fitted a straight line to obtain a new orbital ephemeris, finding $\chir = 2.00$ with one obvious outlier. After the rejection of the offending point, which is the first timing from \citet{Szabo+10aa}, we obtained:
$$ T_0 = {\rm BJD(TDB)} \,\, 2\,455\,176.53878 (27) \, + \, 2.9162383 (22) \times E $$
with $\chir=1.54$.} The bracketed quantities represent the uncertainties in the ephemeris, and have been increased to account for the excess \chir. The full list of transit timings and their references is given in Table\,\ref{tab:minima}. It should be noted that many of the timings are measured from data covering only part of a transit, which is known to reduce their reliability \citep[e.g.][]{Gibson+09apj}.

\begin{table*} \centering \caption{\label{tab:hatp13:final} Final physical properties of the
HAT-P-13 system, compared with results from the literature. Where two errorbars are given,
the first refers to the statistical uncertainties and the second to the systematic errors.}
\begin{tabular}{l r@{\,$\pm$\,}c@{\,$\pm$\,}l c c}
\hline
\ & \mcc{This work (final)} & \citet{Bakos+09apj2} & \citet{Winn+10apj2} \\
\hline
$M_{\rm A}$    (\Msun) & 1.320    & 0.048    & 0.039    & \er{1.219}{0.050}{0.099}    & \er{1.22}{0.05}{0.10} \\
$R_{\rm A}$    (\Rsun) & 1.756    & 0.043    & 0.017    & $1.559 \pm 0.082$           & $1.559 \pm 0.080$     \\
$\log g_{\rm A}$ (cgs) & 4.070    & 0.020    & 0.004    & $4.13 \pm 0.04$             &                       \\
$\rho_{\rm A}$ (\psun) & \mcc{$0.244 \pm 0.013$}        &                             &                       \\[2pt]
$M_{\rm b}$    (\Mjup) & 0.906    & 0.024    & 0.018    & \er{0.853}{0.029}{0.046}    & $0.851 \pm 0.038$     \\
$R_{\rm b}$    (\Rjup) & 1.487    & 0.038    & 0.015    & $1.281 \pm 0.079$           & $1.272 \pm 0.065$     \\
$g_{\rm b}$      (\ms) & \mcc{$10.15 \pm  0.43$}        & $12.9 \pm 1.5$              &                       \\
$\rho_{\rm b}$ (\pjup) & 0.257    & 0.017    & 0.003    & \er{0.375}{0.078}{0.052}    &                       \\[2pt]
\Teq\              (K) & \mcc{$1725 \pm   31$}          & $1653 \pm 45$               &                       \\
\safronov\             & 0.0404   & 0.0023   & 0.0004   & $0.046 \pm 0.003$           &                       \\
$a$               (AU) & 0.04383  & 0.00053  & 0.00043  & \er{0.0427}{0.0006}{0.0012} &                       \\
Age              (Gyr) & \ermcc{3.5}{1.1}{2.9}{0.3}{0.7}& \er{5.0}{2.5}{0.8}          &                       \\
\hline \end{tabular} \end{table*}

The two timings from \citet{Szabo+10aa} both deviate from the orbital ephemeris above, being later by 8.6$\sigma$ and 3.2$\sigma$. Our own analyses of these data return timings which are similarly distant from expectations. A detailed reanalysis of the corresponding data performed by \citet{Fulton+11aj} resulted in a timing for the second of these transits which conflicts less with a linear ephemeris (1.6$\sigma$). The discrepancy of the first transit remains unexplained. A few of the amateur timings are late by a similar amount to this one, but with much larger errorbars.

We conclude that the available data do not provide a clear indication of the existence of TTVs, primarily on the basis that we cannot conceive of a reasonable TTV function which is a significant improvement over a linear ephemeris. The transits which occur later than predicted by our ephemeris are not grouped together, but are interleaved with ones which happen at the expected times. An explanation involving TTVs therefore would require a highly contrived functional form.

So where did the previous suggestions of TTVs come from? \citet{Pal+11mn} used an earlier ephemeris, tuned on the \citet{Bakos+09apj2} and \citet{Szabo+10aa} observations, to show that their transit timings were earlier than expected. The dotted line in Fig.\,\ref{fig:minima} represents this ephemeris and shows that it fails to match the more recent transit timings. \citet{Nascimbeni+11aa2} suggested that a sinusoidal TTV of amplitude 0.005\,d and period 1150\,d was in good correspondance with the observations, as demonstrated by their fig.\,2. We have endeavoured to place this periodic variation, whose parameters were not fully specified, onto Fig.\,\ref{fig:minima}, with a little manual fine-tuning. The 2009-10 transit timings obtained by \citet{Fulton+11aj} clearly dismiss the sinusoidal TTV proposed by \citet{Nascimbeni+11aa2}, leaving a linear ephemeris as the only reasonable option.


\section{Physical properties of the HAT-P-13 system}                                                                        \label{sec:properties}

Several sets of information exist from which the properties of the HAT-P-13\,A,b system can be derived. Analysis of the available light curves has given values for \Porb, $i$, $r_{\rm A}$ and $r_{\rm b}$. The high-precision radial velocities procured by \citet{Bakos+09apj2} and \citet{Winn+10apj2} supply measurements of the velocity amplitude of the star ($K_{\rm A} = 106.04 \pm 0.73$\ms) and the orbital eccentricity ($e = 0.0133 \pm 0.0041$). Analysis of the spectra by \citet{Bakos+09apj2} furthermore lead to estimates of the stellar effective temperature ($\Teff = 5653 \pm 90$\,K) and metallicity ($\FeH = +0.41 \pm 0.08$). Finally, constraints on the properties of the star can be obtained by interpolation within tabulated predictions from stellar evolutionary models.

Our solution process \citep{Me09mn} consists of finding the best agreement between the observed and model-predicted \Teff s, and the measured $r_{\rm A}$ and calculated $\frac{R_{\rm A}}{a}$. This is done using the velocity amplitude of the planet, $K_{\rm b}$, as a solution control parameter and calculating the full system properties using standard formulae \citep[e.g.][]{Hilditch01book}. The system properties comprise the mass, radius, surface gravity and mean density for the star ($M_{\rm A}$, $R_{\rm A}$, $\log g_{\rm A}$ and $\rho_{\rm A}$) and planet ($M_{\rm b}$, $R_{\rm b}$, $g_{\rm b}$ and $\rho_{\rm b}$), the orbital semimajor axis ($a$), the planetary equilibrium temperature and \citet{Safronov72} number (\Teq, \safronov) and an estimate of the evolutionary age of the star.

The statistical errors on the resulting values are calculated using a perturbation analysis \citep{Me++05aa} which yields a full error budget for each output quantity. The use of theoretical models incurs a dependence on stellar theory which is assessed by running separate solutions with each of five different sets of model tabulations (see \citealt{Me10mn} for details). Finally, an alternative empirical estimate of the physical properties is obtained using a calibration \citep{Me11mn} based on eclipsing binary star systems, inspired by \citet{Enoch+10aa} and \citet{Torres++10aarv}. The control parameter $K_{\rm b}$ represents the constraints obtained from stellar theory. Note that the values $g_{\rm b}$, $\rho_{\rm A}$ and \Teq\ are not reliant on constraints from stellar theory \citep{Me++07mn,SeagerMallen03apj,Me11mn}.

The sets of physical properties arising from each of the five stellar model tabulations and from the empirical calibration are given in Table\,A11. For the final results in Table\,\ref{tab:hatp13:final} we adopted the unweighted mean of each parameter from the solutions for the five stellar models. The statistical error is the largest of the individal errorbars, and the systematic error is the standard deviation of the five values. Compared to previous studies, we find a larger and more evolved star and a correspondingly slightly more massive but significantly bigger and hotter planet. The extensive photometric dataset considered in this work leads to more precise measurements of the physical properties, but the uncertainties in $r_{\rm A}$ and $r_{\rm b}$ continue to dominate the error budget. The uncertainty in $M_{\rm A}$ stems mainly from that in \FeH, suggesting that a new spectral synthesis study would also be useful in improving our understanding of the HAT-P-13 system.


\section{Summary}                                                                                                                 \label{sec:summary}

The HAT-P-13 system is unusual in that a transiting hot Jupiter and its host star are accompanied by a clearly-detected third component on a wider orbit. Such a configuration should result in HAT-P-13\,c inducing TTVs within the HAT-P-13\,A,b system, which may be detectable within a comparatively short time period. This possibility has generated substantial interest, resulting in a large body of photometric observations covering many transits of the star by the inner planet. We have assembled the available transit timing measurements and shown that they are most easily explained by a linear ephemeris, albeit with a small number of values which occur later than expected. The discrepant measurements are not clumped together, so could only be explained via highly complex functional forms. Previous claims of TTVs can be attributed to small-number statistics, although continued photometric monitoring has a good chance of turning up something interesting in the future.

We have presented new observations of four transits, obtained using telescope defocussing techniques. Including previously published data, we have ten good sets of transit light curves. These were each analysed within the context of our {\it Homogeneous Studies} project \citep{Me08mn,Me09mn,Me10mn,Me11mn}, and a good agreement between the results was found. We combined them with the measured spectroscopic properties of the host star and several sets of theoretical stellar model predictions to find the physical properties of the system. HAT-P-13 is now well-characterised, although additional photometric and spectroscopic measurements would allow further improvement. We have included it in the TEPCat catalogue\footnote{The {\bf T}ransiting {\bf E}xtrasolar {\bf P}lanets {\bf Cat}alogue can be found online at {\tt http://www.astro.keele.ac.uk/$\sim$jkt/tepcat/}} of the physical properties of transiting planetary systems.

We find a significantly different set of physical properties compared to previous studies, which had access to only two of the ten photometric datasets used here. The star is more massive, larger and more evolved. The planet, whose properties are measured relative to its host star, is similarly heavier and bigger. Its lower density and higher equilibrium temperature place it firmly in the `pM' class advocated by \citet{Fortney+08apj}. Its radius is too large to match the values predicted by the models of \citet{Fortney++07apj} or \citet{Baraffe++08aa}.

\citet{Laughlin++11apj} found that the radius anomaly (the measured radius of a TEP versus that predicted by theoretical models) is correlated with equilibrium temperature, and possibly inversely correlated with host star \FeH. The large radius anomaly and high equilibrium temperature of HAT-P-13\,b corroborate the former observation, but the highly metal-rich nature of the parent star ($\FeH = 0.41 \pm 0.08$) is contrary to the latter suggestion.



\section*{Acknowledgments}

The reduced light curves presented in this work will be made available at the CDS ({\tt http://cdsweb.u-strasbg.fr/}) and at {\tt http://www.astro.keele.ac.uk/$\sim$jkt/}. This observational campaign has been possible thanks to the generous allocation of telescope time by the TAC of the Bologna Observatory and to the invaluable help of the technical staff. JS acknowledges financial support from STFC in the form of an Advanced Fellowship. We thank Andras P\'al for supplying photometric data and the anonymous referee for insightful comments. The following internet-based resources were used in research for this paper: the ESO Digitized Sky Survey; the NASA Astrophysics Data System; the SIMBAD database operated at CDS, Strasbourg, France; and the ar$\chi$iv scientific paper preprint service operated by Cornell University.

\bibliographystyle{mn_new}

\end{document}